\documentstyle [12pt] {article}
\title {CAN CLASSICAL LINEAR HARMONIC OSCILLATOR HOLD
CHAOTIC (FRACTAL) DYNAMICS}

\author {Vladan Pankovi\'c$^\ast$\\ 
$^\ast$Department of Physics, Faculty of Sciences,\\
21000 Novi Sad, Trg Dositeja Obradovi\'ca 4., Serbia}
\begin {document}
\maketitle

\vspace {1.5cm}
\vspace {1.5cm}

\begin {abstract}
In this work a classical linear harmonic oscillator, evolving during a small time interval (so that simple non-linear, second order Taylor approximation of the dynamics is satisfied) and restarting (by a mechanism) in a strictly chosen series of the time moments, is considered. It is shown that given linear oscillator behaves dynamically analogously to discrete logistic equation, which, as an especial case, includes chaotic (fractal) behaviour too. (All this refers too on the analogous quantum systems, e.g. ammonia molecule with simple vibration dynamics of the atoms many times perturbed by measurements in strictly defined time moments.)
\end {abstract}

\section {Introduction}
As it is well-known [1] first-order, discrete difference logistic equation holds extremely simple non-linear form
\begin {equation}
  \frac {\Delta x_{n}}{\Delta t} = {\it a}x_{n-1}(1 - \frac { x_{n-1}}{r}),
  \qquad\mbox{for } n=1, 2, 3, ... .
\end {equation}
Here $x_{n}< r$ represents the population discrete variable and $\Delta x_{n}= x_{n} - x_{n-1}$ its discrete change in time unit $\Delta t$ for $n=1, 2, ... $, while $x_{0}>0$, ${\it a}>0$, and $r>1$ represent the initial population, population growth parameter and carrying capacity parameter.

Simple solution of this equation holds form
\begin {equation}
  x_{n} = (1 + {\it a}\Delta t )
          [\frac {r(1 + {\it a}\Delta t)}{{\it a}\Delta t}]
           \frac {x_{n-1} }{[\frac {r(1 + {\it a}\Delta t)}{{\it a}\Delta t}]}
            (1 - \frac {x_{n-1} }{[\frac {r(1 + {\it a}\Delta t)}{{\it a}\Delta t}]}) 
\end {equation}
That implies
\begin {equation}
  \theta_{n}= \alpha \theta_{n-1}(1-\theta_{n-1}),
  \qquad\mbox{for } n=1, 2, 3, ...            .
\end {equation}
Here 
\begin {equation}
  \alpha = (1 + {\it a}\Delta t ) > 1
\end {equation}
represents the relative population growth parameter, while
\begin {equation}
    \theta_{n} =  \frac {x_{n}}{[\frac {r(1 + {\it a}\Delta t)}{{\it a}\Delta t}]} < 1,
    \qquad \mbox{for } n=0, 1, 2, 3, ...
\end {equation}
represents the relative population discrete variable with initial value $\theta_{0}<1$.
 
As it is well-known too [1], solution, i.e. series (3) for different values of $\alpha$ (4) behaves very differently. For $\alpha$  smaller than 3 (3) converges. For $\alpha$ equivalent to 3 (3) oscillates between two values (bifurcation). For $\alpha$  larger than 3 but smaller than (approximately) 3.57 number of the  stable points around which (3) oscillates grows up 8. For $\alpha$  equivalent to 3.57 any oscillation stops. For $\alpha$  larger and larger than 3.57, but smaller than 4, series (3) behaves larger and larger chaotically (fractally). Finally, for $\alpha$  equivalent and larger than 4 (3) behaves completely chaotically (fractally).

It has been observed and pointed out [1] that extremely rich spectrum of the dynamical behaviour of the discrete logistic equation solution is principally different from extremely pure spectrum of the dynamical behaviour of linear systems, e.g. classical linear harmonic oscillator.

In this work a classical linear harmonic oscillator, evolving during a small time interval (so that simplest non-linear, second order Taylor approximation of the dynamics is satisfied) and restarting (by a mechanism) in a strictly chosen series of the time moments, will be considered. It will be shown that given linear oscillator bechaves analogously to discrete logistic equation, which, as an especial case, includes chaotic (fractal) behaviour too. (All this refers too on the analogous quantum systems, e.g. ammonia molecule with simple vibration dynamics of the atoms [2] many times perturbed by measurements in strictly defined time moments.) 

\section {Classical linear harmonic oscillator with chaotic (fractal) dynamics}
As it is well-known dynamics of a linear harmonic oscillator with coordinate $x$, mass $m$ and elasticity coefficient $k$ is described by simple, second-order, differential Newtonian equation
\begin {equation}
  m\frac {d^{2}x}{dt^{2}} = -kx       .
\end {equation}

Simple solution of given equation is
\begin {equation}
   x = x_{0}\cos \omega t + \frac {v_{0}}{\omega}\sin \omega t       
\end {equation}
where $\omega = (\frac {k}{m})^{\frac {1}{2}}$ represents the ciruclar frequency, $x_{0}$ - initial coordinate  and $v_{0}$ - initial speed.

For sufficiently small $t$, (7) can be approximated by the following second-order Taylor expansion
\begin {eqnarray}
   x &=& x_{0}(1 - \frac {1}{2}\omega^{2}t^{2}) + v_{0}t = x_{0}  +  v_{0}t - \frac {1}{2}\omega^{2}t^{2},\nonumber\\[0mm]
     &=& x_{0} + [\frac {2v^{2}_{0}}{x_{0}\omega^{2}}]
                   \frac {t}{[\frac {2 v_{0}}{x_{0}\omega^{2}}]}
                   (1-\frac {t}{[\frac {2 v_{0}}{x_{0}\omega^{2}}]}),\nonumber\\[0mm]
     &=& x_{0} + \beta \epsilon (1- \epsilon)
\end {eqnarray}
that yields
\begin {equation}
  x - x_{0} = \beta \epsilon (1- \epsilon)                            .
\end {equation}
Here 
\begin {equation}
 \beta = [\frac {2 v^{2}_{0}}{x_{0}\omega^{2}}]
\end {equation}
represents the coordinate growth parameter, while
\begin {equation}
  \epsilon = \frac { t}{[\frac {2 v_{0}}{x_{0}\omega^{2}}]}                   
\end {equation}
represents the relative time moment.

Chose a small (that admits approximation (8)) time moment $\tau_{1}$   and, according to (9), (11), define coordinates difference 
\begin {equation}
    x_{1} - x_{0} = \beta \epsilon_{1} (1- \epsilon_{1})                            
\end {equation}
where 
\begin {equation}
 \epsilon_{1} = \frac {\tau_{1}}{[\frac {2 v_{0}}{x_{0}\omega^{2}}]}            .           
\end {equation}

Suppose that in given time moment $\tau_{1}$ a restarting mechanism returns practically instantaneously linear harmonic oscillator in the initial state (with initial coordinate $x_{0}$ and initial speed $v_{0}$) when given oscillator becomes again to evolve according to (7).

Define new relative time moment
\begin {equation}
   \epsilon _{2} = \beta \epsilon_{1}(1- \epsilon_{1})
\end {equation}
and absolute time moment
\begin {equation}
   \tau_{2} = \epsilon _{2}[\frac {2 v_{0}}{x_{0}\omega^{2}}]            .
\end {equation}

Suppose that in this absolute time moment,
when linear harmonic oscillator holds coordinates difference
\begin {equation}
  x_{2} - x_{0} = \beta \epsilon _{2}(1- \epsilon _{2})      ,
\end {equation}
mentioned restarting mechanism returns practically instantaneously given oscillator in the initial state (with initial coordinate $x_{0}$ and initial speed $v_{0}$). Then this oscillator becomes again to evolve according to (7), etc.

By simple induction relative time moments
\begin {equation}
   \epsilon_{n}= \beta \epsilon_{n-1}(1- \epsilon_{n-1}),
   \qquad\mbox{for } n= 2, 3, ...
\end {equation}
and absolute time moments
\begin {equation}
   \tau_{n} = \epsilon _{n}[\frac {2 v_{0}}{x_{0}\omega^{2}}],
   \qquad\mbox{for } n= 2, 3, ...        
\end {equation}
can be consistently defined. Also, it can be supposed that in any of given absolute time moments (18), when linear harmonic oscillator holds coordinates difference
\begin {equation}
  x_{n} - x_{0} = \beta \epsilon_{n}(1- \epsilon_{n}),
   \qquad\mbox{for } n= 2, 3, ...  ,
\end {equation}
mentioned restarting mechanism returns practically instantaneously given oscillator in the initial state (with initial coordinate $x_{0}$ and initial speed $v_{0}$). Then this oscillator becomes again to evolve according to (7), etc.

As it is not hard to see given classical linear harmonic oscillator (evolving during a small time interval according to (8) and restarting by some mechanism in a strictly chosen series of the time moments (18)) behaves (19) analogously to discrete logistic equation (3). This analogy simply implies that this classical linear harmonic oscillator holds chaotic (fractal) dynamics too when coordinate growth parameter (10) is larger than, approximately, 3.57 for appropriately chosen initial coordinate, initial speed and circular frequency.

\section {Quantum vibration systems (similar to classical linear harmonic oscillator)
 with chaotic (fractal) dynamics}
Feynman [3] suggested a simple example where vibration of the ammonia, $NH_{3}$, molecule is presented as a quantum oscillator in basic aspects similar to classical oscillating systems. Vibration quantum dynamical state of the ammonia molecule in a small time interval $[0, t]$ (that admits second order Taylor expansion approximation) is given by expression
\begin {eqnarray}
    |\psi\rangle &=& \cos \omega t \,|1\rangle + \sin \omega t \,|2\rangle,\nonumber\\
    &\simeq& (1 - \frac {1}{2}\omega^{2}t^{2}]) \,|1\rangle + \omega t \,|2\rangle,
   \qquad\mbox{for } \omega t \ll 1   .
\end {eqnarray}
Here $|1\rangle$ represents state of $N$ atom ``up'' $H$ atoms triangle, $|2\rangle$ - state of $N$ atom ``down'' $H$ atoms triangle, while $\omega = \frac {2\pi}{T}$ represents the circular frequency. 

Suppose that in a small time moment $\tau_1$ measurement of a quantum state $|s\rangle={\it a}\,|1\rangle+{\it b}\,|2\rangle$ (for ${\it a}^{2}+ {\it b}^{2} = 1$ and ${\it a}>{\it b}$) in $|\psi\rangle$ (20) is realized. Then, according to standard quantum mechanical formalism probability of the detection of $|s\rangle$ in $|\psi\rangle$ equals
\begin {eqnarray}
   P_{s1} &=& |\langle s|\psi\rangle|^{2} = {\it a}^{2} + 2{\it a}{\it b}\omega \tau_{1} - ({\it a}^{2}- {\it b}^{2})\omega^{2}\tau^{2}_{1},\nonumber\\
   &=& {\it a}^{2} + [ \frac {(2{\it a}{\it b})^{2}}{({\it a}^{2} - {\it b}^{2})\omega}]  \frac {\tau_{1}}{[ \frac {2{\it a}{\it b}}{({\it a}^{2} - {\it b}^{2})\omega}]} (1 -\frac {\tau_{1}}{[ \frac {2{\it a}{\it b}}{({\it a}^{2} - {\it b}^{2})\omega}]}),\nonumber\\
   &=& {\it a}^{2} + \gamma   \eta_{1} (1- \eta_{1}) .
\end {eqnarray}
Here 
\begin {equation}
 \gamma  = [ \frac {(2{\it a}{\it b})^{2}}{({\it a}^{2} - {\it b}^{2})\omega}] 
\end {equation}
represents the probability growth parameter, while
\begin {equation}
 \eta_{1}= \frac {\tau_{1}}{[ \frac {2{\it a}{\it b}}{({\it a}^{2} - {\it b}^{2})\omega}]}                     
\end {equation}
represents the relative time moment.

Suppose that by detection state $|s\rangle$ be realized. Since exact quantum dynamics of the system (20) is cyclic, $|s\rangle$ will evolve and in some time moment $t_{s}$ it will turn out in the state identical to $|\psi (0)\rangle$ (20). Starting from obtained state (20) as new initial state and the moment of the realization of this state as new initial time moment, we can define relative time moment
\begin {equation}
 \eta_{2} = \gamma  \eta_{1}(1-\eta_{1}) 
\end {equation}
and corresponding absolute time moment 
\begin {equation}
 \tau_{2} = [ \frac {2{\it a}{\it b}}{({\it a}^{2} - {\it b}^{2})\omega}] \eta_{2}   
\end {equation}
in which detection of $|s\rangle$ in $|\psi (\tau_{2})\rangle$ (20) can be again realized. Then, probability of the appearance of $|s\rangle$ by detection equals
\begin {equation}
   P_{s2} = {\it a}^{2} + \gamma  \eta_{2}(1-\eta_{2}) .
\end {equation}

If by detection state $|s\rangle$ be realized, it can simply evolve and in the time moment $t_s$ it will turn out in the state identical to $|\psi (0)\rangle$ (20), etc.

By simple induction we can define a series of the relative and absolute time moments
\begin {equation}
 \eta_{n}= \gamma  \eta_{n-1}(1-\eta_{n-1}),
 \qquad\mbox{for } n= 2, 3, ...        
\end {equation}
\begin {equation}
 \tau_{n} = [ \frac {2{\it a}{\it b}}{({\it a}^{2} - {\it b}^{2})\omega}] \eta_{n},
 \qquad\mbox{for }  n= 2, 3, ...        
\end {equation}
as well as probabilities that $|s\rangle$ be detected in $|\psi (\tau_{n})\rangle$ (20)  for  $n= 2, 3, ...$        
\begin {equation}
   P_{sn} = {\it a}^{2} +  \gamma  \eta_{n}(1-\eta_{n}),
 \qquad\mbox{for }  n= 2, 3, ...        .
\end {equation}

According to (29) probabilities differences can be defined by expression
\begin {equation}
   P_{sn} - {\it a}^{2} =  \gamma  \eta_{n}(1-\eta_{n}),
 \qquad\mbox{for }  n= 2, 3, ...        .
\end {equation}

As it is not hard to see given quantum vibrator, evolving during a small time interval according to (20) and many times restarting by $|s\rangle$ detection in strictly determined time moments and cyclic evolution,  behaves analogously to classical linear harmonic oscillator discussed in the previous section of this work.  In other words, given quantum vibrator holds, for strictly chosen series of the discrete detection time moments (28), probabilities differences that behave analogously to discrete differences of the linear harmonic oscillator coordinates (19), or solution of discrete logistic equation (3). This analogy simply implies that this given quantum vibrator holds chaotic (fractal) dynamics of the probabilities differences when probability growth parameter (22) is larger than, approximately, 3.57 for appropriately chosen {\it a}.
 
\section {Conclusion}
In conclusion we can repeat and point out the following. In this work a classical linear harmonic oscillator, evolving during a small time interval (so that simple non-linear, second order Taylor approximation of the dynamics is satisfied) and many times restarting (by a mechanism) in a strictly chosen series of the time moments, is considered. It is shown that given linear oscillator behaves analogously to discrete logistic equation, which, as an especial case, includes chaotic (fractal) behaviour too. (All this refers too on the analogous quantum systems, e.g. ammonia molecule with simple vibration dynamics of the atoms many times perturbed by measurements in strictly defined time moments.)

\section {References}

\begin {itemize}

\item [[1]] R. M. May, Nature {\bf 261} (1976) 459
\item [[2]] R. P. Feynman, R. B. Leighton, M. Sands, {\it The Feynman Lectures on Physics, Vol. 3} (Addison-Wesley Inc., Reading, Mass. 1963)

\end {itemize}

\end {document}